%% file: ms.tex
\documentclass[usenatbib]{mn2e}
\usepackage{natbibmnfix,graphicx,times, amsmath}

\include{defns}

\newcommand{\DC}{{\rm DC}}
\newcommand{\taueff}{\tau^{\rm eff}}
\newcommand{\taueffa}{\tau^{\rm eff}_{\alpha}}

\newcommand{\meanG}{\langle\Gamma_{12}\rangle}

\def\myputfigure#1#2#3#4#5%
{\vskip#5pt\makebox[0pt]{\hskip#2in
\includegraphics[width=#3\textwidth]{#1}}\vskip#4pt\hfill}

\newcommand\lsim{\mathrel{\rlap{\lower4pt\hbox{\hskip1pt$\sim$}}
        \raise1pt\hbox{$<$}}}
\newcommand\gsim{\mathrel{\rlap{\lower4pt\hbox{\hskip1pt$\sim$}}
        \raise1pt\hbox{$>$}}}

\begin{document}

%\submitted{Submitted to the MNRAS}

\title[The Ionizing Background Following Reionization]{The Inhomogeneous Ionizing Background Following Reionization}

\author[Mesinger \& Furlanetto]{Andrei Mesinger$^1$\thanks{Hubble Fellow; email: mesinger@astro.princeton.edu} \& Steven Furlanetto$^2$ \\
$^1$Department of Astrophysical Sciences, Princeton University, Princeton, NJ 08544, USA \\
$^2$Department of Physics and Astronomy, University of California, Los Angeles, CA 90095, USA}
\voffset-.6in

\maketitle

\begin{abstract}
We study the spatial fluctuations in the hydrogen ionizing background in the epoch following reionization ($z\sim$ 5--6).   The rapid decrease with redshift in the photon mean free path (m.f.p.), combined with the clustering of increasingly rare ionizing sources, can result in a very inhomogenous ionizing background during this epoch.  We systematically investigate the probability density functions (PDFs) and power spectra of ionizing flux, by varying several parameters such as the m.f.p., minimum halo mass capable of hosting stars, and halo duty cycle.  In order to be versatile, we make use of analytic, semi-numeric and numeric approaches.  Our models show that the ionizing background indeed has sizable fluctuations during this epoch, with the PDFs being a factor of few wide at half of the maximum likelihood.  The clustering of sources dominates the width of the PDFs, so analytic models must take large-scale clustering into account. The distributions also show marked asymmetries, with a high-value tail set by clustering on small scales, and a shorter low-value tail which is set by the mean free path. The power spectrum of the ionizing background is much more sensitive to source properties than the PDF and can be well-understood analytically with a framework similar to the halo model (usually used to describe dark matter clustering).   Nevertheless, we find that \lya\ forest spectra are extremely insensitive to the details of the UVB, despite marked differences in the PDFs and power spectra of our various ionizing backgrounds. Assuming a uniform ionizing background only underestimates the value of the mean ionization rate, $\langle\Gamma_{12}\rangle$, inferred from the \lya\ forest by a few percent.  Instead, analysis of the \lya\ forest is dominated by the uncertainties in the density field.  Thus, our results justify the common assumption of a uniform ionizing background in \lya\ forest analysis even during this epoch.
\end{abstract}


\section{Introduction}
\label{sec:intro}

The metagalactic hydrogen-ionizing ultraviolet background (UVB) is probably the most influential, pervasive radiation field throughout most of cosmic history.  Much investigation has gone into understanding its properties and its impact on the intergalactic medium (IGM) (see \citealt{Meiksin07} for a recent review).  At low redshifts ($z\lsim$ 3--4), the ionizing photon mean free path (m.f.p.) is very large, and so each gas element effectively ``sees'' many sources of ionizing radiation.  A uniform UVB seems to be an accurate approximation in this regime, with spatial fluctuations contributing only a few percent of the mean \citep{Zuo92, FS93, GH02, MW04, Croft04, Bolton06}.

However, the m.f.p. decreases rapidly with redshift, with extrapolations predicting that it should approach values of $\lmfp\sim$ 30--40 comoving Mpc by $z\sim$ 5--6 (e.g. \citealt{Storrie-Lombardi94, Miralda-Escude03}). Such modest values of the m.f.p. at $z\gsim5$ can induce significant fluctuations in the UVB, sourced by cosmic variance in the number of ionizing sources inside each $\lmfp$ horizon.  Additionally, the 1/$r^2$ flux attenuation, combined with the clustering of sources, can introduce additional scatter in the UVB on scales smaller than $\lmfp$ \citep{Zuo92, MW04}.  These fluctuations have been largely ignored in high-redshift analyses (though see \S 5 in \citealt{BH07} and \S 4 in \citealt{Feng08}).

Here we undertake a systematic study of the post-reionization metagalactic ionizing background at $z\sim$ 5--6.  We vary source prescriptions and $\lmfp$ in an attempt to {\it infer general conclusions about the theoretical properties and observational importance of spatial fluctuations in the UVB}.  We focus on the simplest statistics, the probability density function (PDF) and power spectrum, which are most likely to be observed.  Given the dearth of knowledge concerning the properties of absorbers and sources at high-$z$, we make use of a range of methods that include analytic, semi-numeric and numeric approaches in order to be comprehensive.  We compare simple analytic models of these quantities to the numerical results in order to shed light on the origin and importance of the variations; we will show that analytic models are accurate in some, but not all, cases.

This paper is organized as follows.  In \S \ref{sec:flux}, we outline the various modeling tools we use to study the fluctuations in the ionizing background.  In \S \ref{sec:PDF} and \S \ref{sec:ps} we present the PDFs and power spectra of various UVBs, respectively.  In \S \ref{sec:forest}, we study the impact of a spatially varying UVB on the Ly$\alpha$ forest.  Finally, in \S \ref{sec:conc}, we summarize our findings and offer conclusions.

We quote all quantities in comoving units, with the exception of flux, and we denote proper units with a prefix 'p'.  We adopt the background cosmological parameters ($\Omega_\Lambda$, $\Omega_{\rm M}$, $\Omega_b$, $n$, $\sigma_8$, $H_0$) = (0.72, 0.28, 0.046, 0.96, 0.82, 70 km s$^{-1}$ Mpc$^{-1}$), matching the five--year results of the {\it WMAP} satellite \citep{Dunkley08, Komatsu08}.

\section{Generating Inhomogeneous Flux Fields}
\label{sec:flux}

\subsection{Semi-Numerical Simulations with Flux Tool}
\label{sec:semi}

We generate our ionizing flux fields following the procedure described in \citet{MD08}.  We briefly outline the procedure below.

We begin with halo fields at $z=5$ and $z=6$ generated with the semi-numerical simulation DexM\footnote{http://www.astro.princeton.edu/$\sim$mesinger/DexM/}, which has been shown to reproduce the correct number density and clustering properties of halos well into the quasi-linear and non-linear regimes (\citealt{MF07}; Fig. 1 in \citealt{Dijkstra08}; Mesinger et al. in preparation).  Our simulation boxes are 150Mpc on a side, with a 150/1800 Mpc halo grid cell size.  The velocity fields used to perturb the halo field were generated on a lower resolution 900$^3$ grid; thus our final halo field resolution is $\Delta x =$ 150/900 = 0.17 Mpc.

For each halo field, we create a corresponding UV flux field on a 150$^3$ grid (spatial resolution of 1 Mpc), by performing a halo mass/$r^2$ weighted sum.  Specifically, we compute the flux of ionizing photons (in units of ionizing photons s$^{-1}$ pcm$^{-2}$) with

\begin{equation}
\label{eq:sum}
\flux= \frac{(1+z)^2}{4 \pi} \sum_{M_i>\Mmin} X_i(\DC) \frac{M_i\times \epsilon_{\rm ion}}{|{\bf x} - {\bf x_i}|^2} ~ e^{-|{\bf x} - {\bf x_i}|/\lmfp} ~, 
\end{equation}

\noindent  where ${\bf x}$ is the location of the cell of interest, $M_i$ is the total halo mass, ${\bf x_i}$ is its location, and the factor of $(1+z)^2$ converts the factor $|{\bf x} - {\bf x_i}|^2$ from comoving into proper units.  We also include a duty cycle\footnote{Note that there are several definitions of the term ``duty cycle'' in the literature, some of which include this same parameter in the star-formation timescale of each halo (an assumption only valid in extremely idealized scenarios).  We adopt the simple definition above of the duty cycle, namely the fraction of dark matter halos which are ``on'' (i.e. hosting active galaxies emitting ionizing photons into the IGM; c.f. \citealt{Lee08}).}
 parameter through the random variable $X_i(\DC)$, which has a value of 1 with likelihood $\DC$ or 0 with likelihood $1-\DC$.
  Finally, the factor $\epsilon_{\rm ion}$ in eq. (\ref{eq:sum}) denotes the rate at which ionizing photons are released into the IGM by a dark matter halo per unit mass.  
Unless stated otherwise, we assume a fiducial value of $\epsilon_{\rm ion}(z) = 3.8 \times 10^{58}/\DC \left[\frac{\Omega_b}{\Omega_{\rm M}} \frac{1}{t_H(z)}\right]\hs\frac{{\rm photons}}{M_{\odot}\hs {\rm s}}.$, which provides a good fit to the observed luminosity functions of Ly$\alpha$ emitting galaxies (LAEs) \citep{Shimasaku06, Kashikawa06, DWH07, SLE07, McQuinn07LAE}, and the $z=6$ Lyman Break galaxies (LBGs) \citep{Bouwens06}.
However, as we are mostly concerned with the shape of the flux PDF, where applicable we present results in units of the mean flux or intensity, $J/\langle J\rangle$.

In constructing the UVB, we also assume a minimum halo mass able to host stars, $\Mmin$, and explore mainly two different values for this parameter: $1.6\times10^8$ and $1.4\times10^9$ $\Msun$. The former value corresponds to a virial temperature of 10$^4$ K at these redshifts, and represents the regime of ineffective feedback on atomically cooled halos during reionization \citep{MD08, OGT08}.  The later value represents the regime of very inefficient star formation inside such small halos, perhaps due to strong radiative and/or mechanical feedback (e.g. \citealt{Yepes97, Scannapieco06, PS08}).  The exact value of the higher $\Mmin$ was also motivated by the fact that the number density of halos with mass greater than $1.4\times10^9$ $\Msun$ is roughly ten times smaller than the number density of halos with masses greater than $1.6\times10^8$ $\Msun$ at $z\approx5$, thus allowing us to compare results at fixed source number density with values of $\DC=$ 1.0 and 0.1 in the two models, respectively.

%Finally, the units of the flux $\flux$ are 'ionizing photons s$^{-1}$ pcm$^{-2}$'. On the other hand, the ionizing background is commonly quantified in terms of its mean intensity $J(\nu)$ which has the units 'erg s$^{-1}$ pcm$^{-2}$ sr$^{-1}$ Hz$^{-1}$'. If the spectrum of the ionizing background radiation field is assumed to be of the form $J(\nu)=J_{\rm 21}(\nu/\nu_H)^{-\alpha}\times 10^{-21}$ erg s$^{-1}$ Hz$^{-1}$ pcm$^{-2}$ sr$^{-1}$, then the $J_{21}$ value in the cell located at $({\bf x}, z)$ is trivially obtained from $\flux$ through the equation  $J_{\rm 21}({\bf x}, z)= \flux \times h_{\rm p}\alpha/(4\pi) / 10^{-21}\hs{\rm erg}\hs{\rm s}^{-1} \hs{\rm pcm}^{-2}\hs{\rm Hz}^{-1}\hs{\rm sr}^{-1}$.  For quantitative estimates of $J_{\rm 21}$, we use the value of $\alpha\sim5$ in the above conversion which characterizes stellar sources \citep{TW96, BL01}.  Note however that this choice is purely for presentation purposes; in fact we are mostly conserned with the shape of the flux PDF.  Thus, where applicable we present results in units of the mean flux or intensity, $J/\langle J\rangle$.

\subsection{Hydrodynamical Simulations with Flux Tool}
\label{sec:num}

Additionally, we generate a flux field corresponding to the $z=5.71$ source field from the cosmological hydrodynamical simulation presented in \citet{TCL08} (their $z=6$ reionization model).  This simulation is fixed-grid, 143 Mpc on a side, and includes prescriptions for modeling dark matter, baryons and ionizing photons (for details see \citealt{TC07} and \citealt{TCL08}).  The density field was calculated on grid of 0.19 Mpc cells, which resolves the Jeans length in the mean density, ionized intergalactic medium (IGM) by a factor of few, and then smoothed to a cell size of 0.74 Mpc.   Each halo's instantaneous star-formation rate (SFR) is proportional to the instantaneous gas accretion rate, and enters our eq. (\ref{eq:sum}) in place of the halo's mass, $M_i$, with the normalization adjusted accordingly so as to match the mean value of the UVB.  We make use of this flux field, combined with the $z=5.71$ gas density fields from the hydro-simulation, to generate more accurate mean Ly$\alpha$ forest flux decrement statistics in \S \ref{sec:forest}.  At these redshifts and scales, our semi-numerically generated density fields somewhat over-predict the rare voids which dominate the this statistic (Mesinger et al., in preparation).  Furthermore, the ray-tracing algorithm from the numerical simulation over-predicts the fluctuations in the flux field, due to an insufficient number of ray splittings (Trac 2008, private communication).  This problem only becomes severe following reionization, but since this is the epoch we study here, we use this hybrid prescription in some of the results below (i.e. we do not use the radiative transfer field from the simulation).

\begin{figure*}
\vspace{+0\baselineskip}
{
\includegraphics[width=0.45\textwidth]{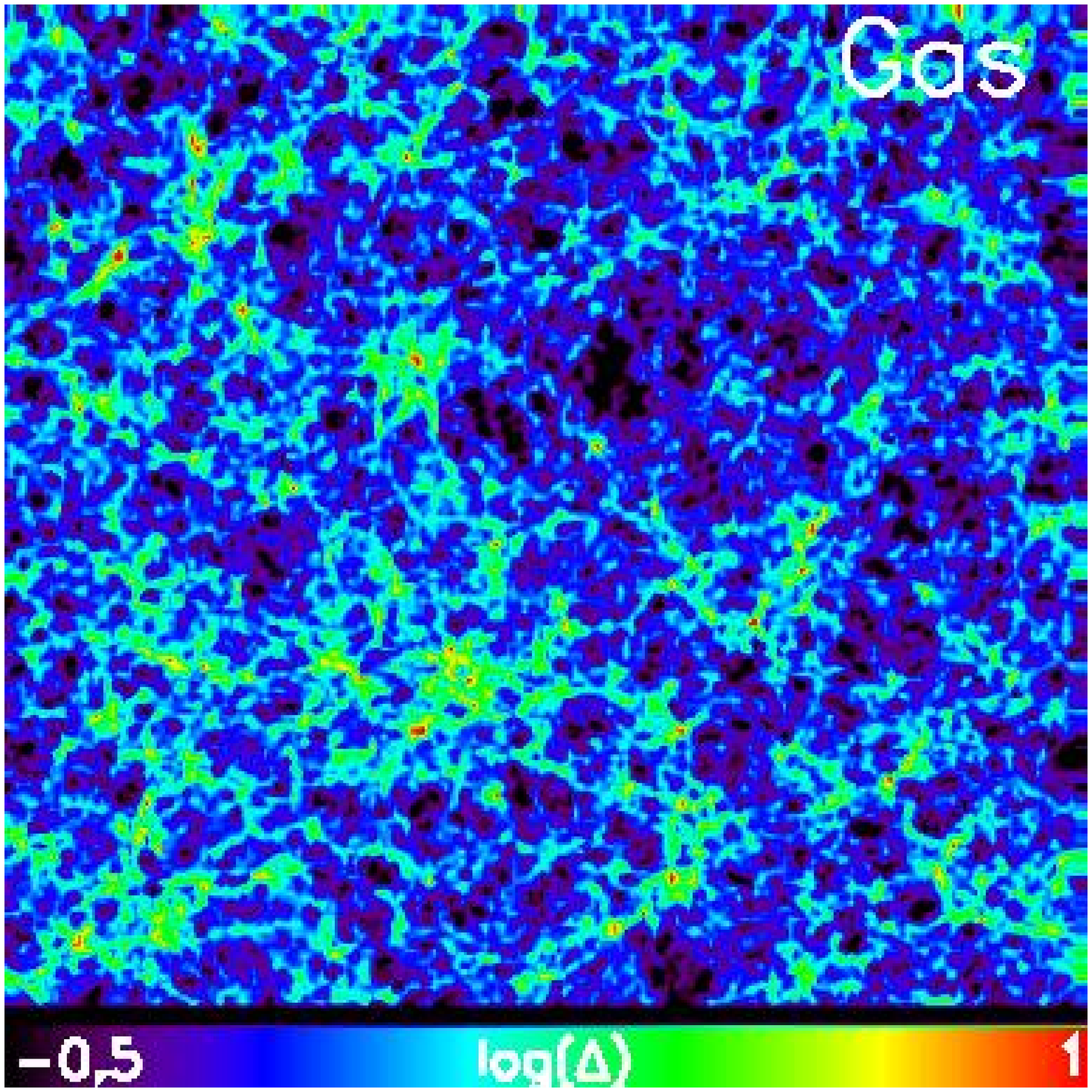}
\includegraphics[width=0.45\textwidth]{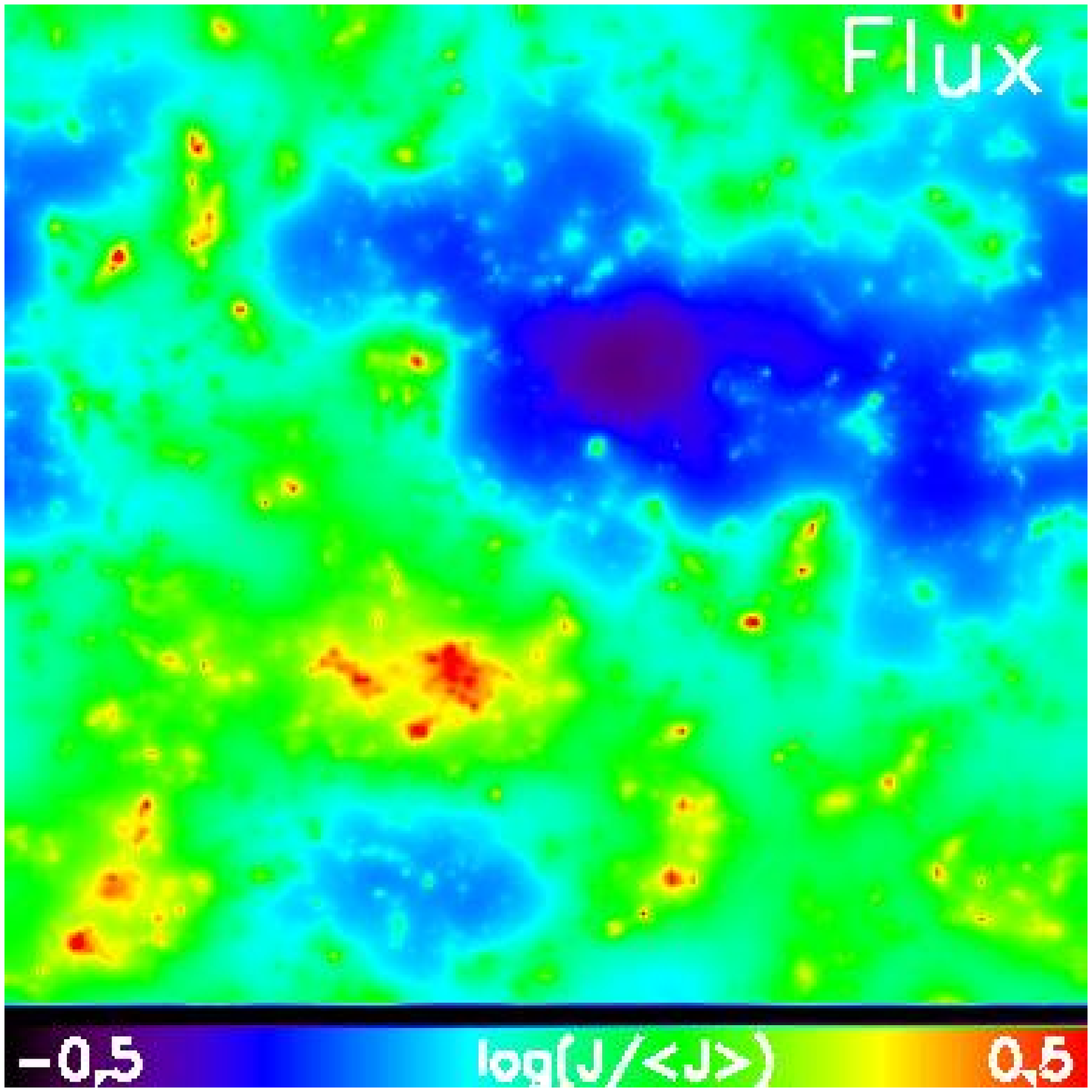}
}
\caption{
Slices through the $z=5.71$ gas density ({\it left}) and UVB ({\it right}) fields using the hydrodynamical simulation of \citet{TCL08} and our simple flux generation tool.  Each slice is 143 Mpc on a side and 0.74 Mpc thick.  The UV flux is generated using the star-formation rates obtained from the simulation and the prescription outlined in \S \ref{sec:flux}, assuming $\lmfp=30$ Mpc, DC=1, and $\Mmin\sim 3 \times 10^8 \Msun$.
\label{fig:pics}
}
\vspace{+1.5\baselineskip}
\end{figure*}

In Fig. \ref{fig:pics}, we present a 0.74 Mpc thick slice through the density and flux fields based on this $z=5.71$ simulation output from \citet{TCL08}.  The flux fields were calculated assuming $\lmfp=30$ Mpc, and $\DC=1$.  Even for such a moderately high choice of $\lmfp$, we can qualitatively see that there is significant inhomogeneity in the flux fields.  Furthermore, it is evident that the flux and density fields are highly correlated, an issue we will return to in \S \ref{sec:forest}.

\subsection{Analytic Model}
\label{sec:anal}

We complement these numerical techniques with a relatively simple analytic model.  While this approach cannot incorporate all of the physics provided by the semi-numeric and numeric models, we will find it to be helpful in elucidating the physics of the background radiation field.  As in the remainder of this paper, we will focus on computing two statistical descriptions of the flux field:  the PDF and the power spectrum.

\subsubsection{Flux PDF}

The PDF of the radiation background $J$, normalized to its mean value $\langle J\rangle$, can be computed exactly for randomly distributed sources, provided that we assume a constant attenuation length $\lambda_{\rm mfp}$ for ionizing photons much smaller than the Hubble length \citep{Zuo92, MW04}.  In this case, the crucial parameter is the number of sources within each attenuation length, $\bar{N}_0 = (4 \pi/3) n_i \lambda_{\rm mfp}^3$, where $n_i$ is the source number density, and
\begin{eqnarray}
{dp (>X|\bar{N}_0) \over dX} & = &  {1 \over \pi} \int_0^{\infty} ds \, \exp \left[ -s \bar{N}_0 \int dx \, x \phi(x) \, {\rm Im} \, G(sx) \right] 
\nonumber \\
& & \times \cos \left[ -sX + s \bar{N}_0 \int dx \, x \phi(x) {\rm Re} \, G(sx) \right],
\label{eq:jdist_r0}
\end{eqnarray}
where $X=J/J_\star$, $\phi(x)$ is the source luminosity function, in units of the mean luminosity $L_\star$ and normalized to have unit integral, $J_\star = L_\star/(4 \pi \lambda_{\rm mfp})^2$,
\begin{equation}
G(t) = \int_0^\infty du \, \tau^3(u) e^{itu},
\label{eq:Gdefn}
\end{equation}
and $u=e^{-\tau}/\tau^2$.  In this limit, the mean background is $\langle X \rangle = 3 \bar{N}_0$, or $\langle J \rangle \propto \bar{N}_0 J_\star \propto L_\star r_0$.  

At the end of reionization, $\lambda_{\rm mfp} \gsim 20$~Mpc and (typically) $\bar{N}_0 \gsim 10^5$.  Thus, the Poisson fluctuations in the galaxy counts are very small and one might expect the ionizing background to be extremely uniform.  However, equation~(\ref{eq:jdist_r0}) assumes that the source counts within each attenuation volume are Poisson distributed around the fixed value $\bar{N}_0$; in actuality, deterministic halo clustering causes fluctuations in the \emph{expected} number of halos within each attenuation volume, with a fractional amplitude $\sim \bar{b} \sigma(\lambda_{\rm mfp})$ where $\bar{b}$ is the mean bias of the star-forming galaxies.  For our fiducial model at $z=6$, we find $\bar{b} \sim 2.5$, which yields fractional variations of order $\sim 2$ for $\lambda_{\rm mfp}=30$~Mpc.  

To crudely account for this deterministic clustering, we approximate the underlying density field as linear on scales $\sim \lambda_{\rm mfp}$ (so that the PDF of density, $p(\delta)$, is a gaussian with variance $\sigma^2(\lambda_{\rm mfp})$ and vary the expected number of sources within each attenuation volume according to linear theory, so that $\bar{N}_0 \rightarrow \bar{N}_0(\delta,\lambda_{\rm mfp})$, which depends on the conditional mass function for a region with a fractional overdensity $\delta$ on a scale $\lambda_{\rm mfp}$ \citep{LC93}.  Thus
\begin{equation}
{dp(>X) \over dX} = \int d\delta \, p(\delta) {dp(X|\bar{N}_0[1+\bar{b}\delta]) \over dX}.
\label{eq:jdistbn_cluster}
\end{equation}
We shall see that this is only a crude approximation, for two reasons.  First, our prescription for the expected source density is not terribly accurate, since the typical fractional fluctuation in the galaxy counts is not too much smaller than unity, and because we effectively assign the modified source density to all neighboring regions as well.  More importantly, nonlinear clustering is actually quite important to the PDF.

\subsubsection{Flux Power Spectrum}

Similar techniques also allow one to compute the correlation function or power spectrum of the intensity field due to randomly distributed sources \citep{Zuo93}.  However, in this case it is possible to include (linear) clustering exactly by using arguments similar to the ``halo" model that describes dark matter and galaxy clustering.

The intensity field around a single galaxy of mass $m$ is
\begin{equation}
J(r|M_i) = {\epsilon_{\rm ion} M_i \exp(-r/\lambda_{\rm mfp}) \over (4 \pi r)^2}.
\label{eq:haloprofile}
\end{equation}
In our calculations, the normalization factor is set as in the semi-numeric model.  The total intensity field is simply the sum of the profiles around each galaxy in the Universe.  By analogy to the halo model, the power spectrum can then be written as the sum of two terms, one arising from correlations within a single source's envelope, and one arising from correlations between the locations of 
two sources \citep{Scherrer91, CooraySheth03}.  These two terms both depend on the normalized Fourier transform of the single source profile, equation~(\ref{eq:haloprofile}):
\begin{equation}
u_J(k) = {\arctan (k \lambda_{\rm mfp}) \over k \lambda_{\rm mfp}},
\end{equation}
which is independent of galaxy mass (because it only describes the shape, not the amplitude, of the intensity profile).  Note that $u_J \rightarrow 1$ for $k \lambda_{\rm mfp} \ll 1$ and $u_J \rightarrow \pi/(2 k \lambda_{\rm mfp})$ for $k \lambda_{\rm mfp} \gg 1$.

The ``single-source" term, describing correlations within randomly distributed sources' profiles, is
\begin{equation}
P_{1s}(k) = |u_J(k)|^2 \int_{\Mmin}^\infty dm \, n(m) \left( {m \over \bar{\rho}_{\rm gal}} \right)^2 ,
\label{eq:onesource}
\end{equation}
where $n(m)$ is the mass function of the sources (or the halo mass function in our calculations) and $\bar{\rho}_{\rm gal}=n_i \bar{m}_{\rm gal}$ is the mean mass density of galaxies (and $\bar{m}_{\rm gal}$ is the mean mass of the ionizing sources).  Note that the single-source correlation function can also be derived from the two-point probability density (\citealt{Zuo93}; see also \citealt{Furlanetto09-heliumcorrs}).   Furthermore, the amplitude is determined by the luminosity function, but the shape is entirely determined by our imposed attenuation profile.

Similarly, the term describing the correlations between two sources may be written
\begin{equation}
P_{2s}(k) = |u_J(k)|^2 \bar{b}^2 P_{\rm lin}(k,z) \left[ \int_{\Mmin}^\infty dm \, n(m) \left({m \over \bar{\rho}_{\rm gal}} \right) \right]^2,
\end{equation}
where $P_{\rm lin}(k,z)$ is the linear matter power spectrum.  Again, this two-point function has a shape fixed by the source profiles, but an amplitude that depends on the nature of the sources.  
In fact, the term in square brackets is simply equal to unity; we have left it in only for comparison to the halo model, where $u(k)$ is a function of $m$ and so must be left inside the integral.
Because $|u_J|^2 \rightarrow 1$ for small $k$, the two-source term traces the linear matter power spectrum on these scales, before being truncated at $k \lambda_{\rm mfp} \sim 1$.

If we take a simple model in which all sources have a fixed mass, we find that
\begin{equation}
{P_{2s} \over P_{1s}} \approx {\bar{b}^2 n_i P_{\rm lin}(k) }.
\end{equation}
Thus the two-source term becomes more and more important as the number density of sources, their bias, and the underlying linear clustering increase:  this is simply because all of these increase the importance of deterministic clustering relative to the correlations inherent to the flux profile itself.

\section{The Flux PDF}
\label{sec:PDF}

\subsection{Attenuation}
\label{sec:att_pdf}

As discussed above, we account for flux attenuation through the factor $e^{-r/\lmfp}/(4\pi r^2)$ in eq. (\ref{eq:sum}).  Thus flux attenuation is taken into account in a spatially homogeneous way through the free parameter $\lmfp$. Although such an approach misses the obvious correlation of absorbers (LLSs and DLAs) with sources, it had several benefits over using the typical approximate radiative transfer algorithms in cosmological hydrodynamical simulations.  Firstly, our approach is versatile, easy to calibrate and to run quickly for large-scale boxes.  This allows us to easily explore the effects of varying $\lmfp$, which is important, given the large uncertainties in this value at these redshifts (e.g. \citealt{Faucher-Giguere08}).
Secondly, our formalism allows for easy comparisons with analytic models, and provides direct insight into the features of the UVB.  Finally, and most importantly, large-scale cosmological hydrodynamical simulations {\it do not directly resolve LLSs and DLAs}; thus they too must employ analytic prescriptions to populate their box with absorbers.
We also note that the mean free paths at these redshifts is so large ($\gsim20$ Mpc; \citealt{Storrie-Lombardi94, MHR99, Miralda-Escude03, Faucher-Giguere08}) that flux received by each cell is sourced by tens of thousands of sources, likely averaging out detailed radiative transfer effects.  We thus do not expect our results to be significantly affected by the fact that we neglect the details of radiative transfer and the clustering of absorbers.

Nevertheless, stochastic fluctuations in the absorber population will help to increase the variance in the radiation field:  if LLSs are primarily responsible for absorption, each attenuation length will contain only a single absorber on average and the absorption length will vary quite strongly.  However, in practice $\ga 50\%$ of the absorption is typically due to lower-column density absorbers (see, e.g., the Appendix to \citealt{FO05}), so this is probably not a large factor for us.

\begin{figure}
\vspace{+0\baselineskip}
\myputfigure{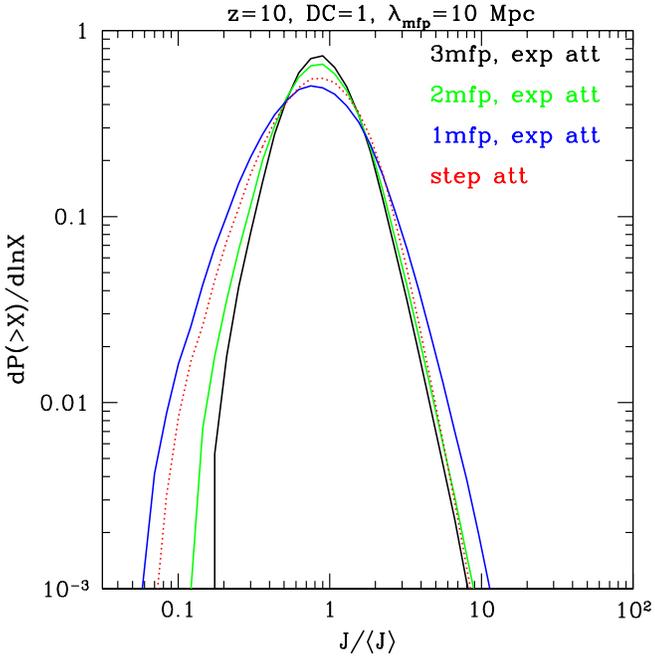}{3.3}{0.5}{.}{0.}
\caption{
Flux PDFs at $z=10$, assuming $\lmfp=10$ Mpc, $\Mmin=10^8 \Msun$ and DC=1.0, calculated according to eq. (\ref{eq:sum}) but truncating the sum over contributing sources at distances of 1$\lmfp$, 2$\lmfp$, and 3$\lmfp$ ({\it left to right solid curves}).  The dotted curve assumes a step-function attenuation, essentially $\tau=0$ for distances $r<\lmfp$ and $\tau=\infty$ at distances $r>\lmfp$.
\label{fig:mfp_att}}
\vspace{-1\baselineskip}
\end{figure}

In Figure \ref{fig:mfp_att}, we plot the flux PDFs for a $L=100$ Mpc box with $0.5$ Mpc cells at $z=10$, $\lmfp=10$ Mpc, and $\Mmin=10^8$ $\Msun$.  We show the effects of truncating the sum over contributing sources in eq. (\ref{eq:sum}) at distances of 1$\lmfp$, 2$\lmfp$, and 3$\lmfp$ ({\it left to right solid curves}).  The distributions appear to converge at 2$\lmfp$ $\rightarrow$ 3$\lmfp$.  
  The high end tails of the distribution are dominated by nearby ionizing sources and are fairly insensitive to the effective horizon (i.e. the high-value tails are roughly constant as more distant sources are included in the UVB calculation).  When plotted in unnormalized units (i.e. $J$ instead of $J/\langle J \rangle$), this overlap is more pronounced and the leftmost (blue) curve shifts to even smaller values. In contrast, the low-end tails of the PDFs correspond to locations far away from ionizing sources and so are very sensitive to the inclusion of more distant sources, which roughly translates to changing the amplitude an effective homogeneous UVB. If sources were homogeneously distributed, truncating the flux distribution at distances greater than $n \lmfp$ would underestimate the ionizing flux by a factor of $(1-e^{-n})$.  The clustering of sources results in the more complicated behavior seen in Fig. \ref{fig:mfp_att}.

The dotted curve in Fig. \ref{fig:mfp_att} neglects the exponential attenuation term in eq. (\ref{eq:sum}), and instead only includes the contributions of sources at distances $r<\lmfp$.  This is analogous to assuming a step-function optical depth so that $\tau=0$ for distances $r<\lmfp$ and $\tau=\infty$ at distances $r>\lmfp$.  Such a prescription can simplify computation in analytic models (e.g., \citealt{Zuo92}) and numerical calculations (e.g., \citealt{Bolton06}) and seems to result in fairly accurate flux distributions, though it slightly overestimates the number of faint regions.  This overestimate worsens somewhat if one integrates sources only out to 1$\lmfp$ but also includes exponential attenuation (e.g. \citealt{Santos08}).
%AM: moved sentance to end here

\begin{figure}
\vspace{+0\baselineskip}
\myputfigure{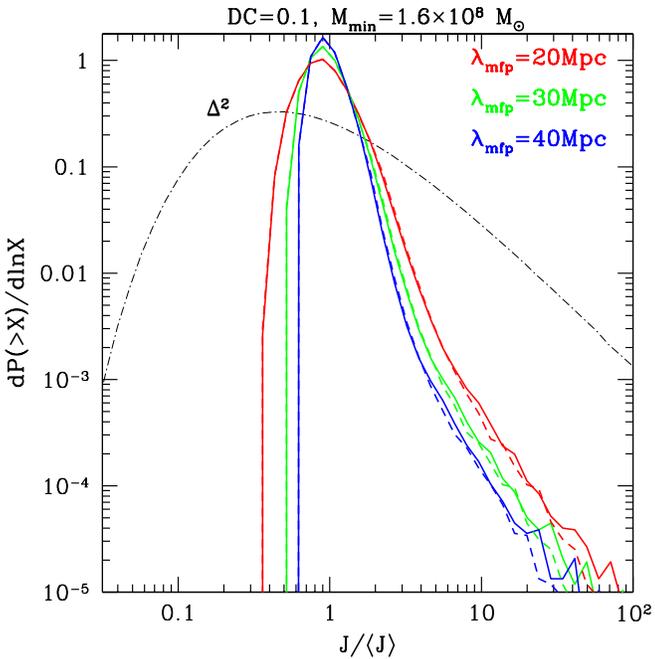}{3.3}{0.5}{.}{0.}
\caption{
Flux PDFs at $z=5$ ({\it solid curves}) and $z=6$ ({\it dashed curves}) generated assuming $\Mmin=1.6\times10^8 \Msun$ and DC=0.1.  Starting from the left side of the figure, curves correspond to values of $\lmfp=$ 20, 30, 40 Mpc, respectively.  For comparison, the dot-dashed curve shows the squared overdensity, $\Delta^2 \equiv (\rho/\bar{\rho})^2$, taken from the hydrodynamical simulation at $z=5.71$ with cell size 0.74 Mpc.
\label{fig:mfp_pdfs}}
\vspace{-1\baselineskip}
\end{figure}

In Figure \ref{fig:mfp_pdfs}, we explore the impact of varying $\lmfp$.  Flux PDFs at $z=5$ ({\it solid curves}) and $z=6$ ({\it dashed curves}) are shown assuming $\Mmin=1.6\times10^8 \Msun$ and DC=0.1.  Note that the evolution of structure from $z=10$ in Fig. \ref{fig:mfp_att} to $z=$ 5 and 6 in Fig. \ref{fig:mfp_pdfs} preferentially suppresses the low-end tail of the PDFs as regions containing few ionizing sources become extremely rare.  This results in narrower and more asymmetric distributions.

 As expected, increasing the m.f.p. also results in narrower flux distributions, as the gas cells see increasingly representative volumes of space inside their horizon.  We note however that the $\lmfp=40$ Mpc distributions are still fairly broad.  Understandably, when plotted in unnormalized units (i.e. $J$ instead of $J/\langle J \rangle$), the high-value tails overlap as they are dominated by nearby halos.  Perhaps surprisingly, we see very little variation of the flux PDF with redshift, over the range $z=$ 5 to 6.

\subsection{Source Clustering}
\label{sec:cluster_pdf}

\begin{figure}
\vspace{+0\baselineskip}
\myputfigure{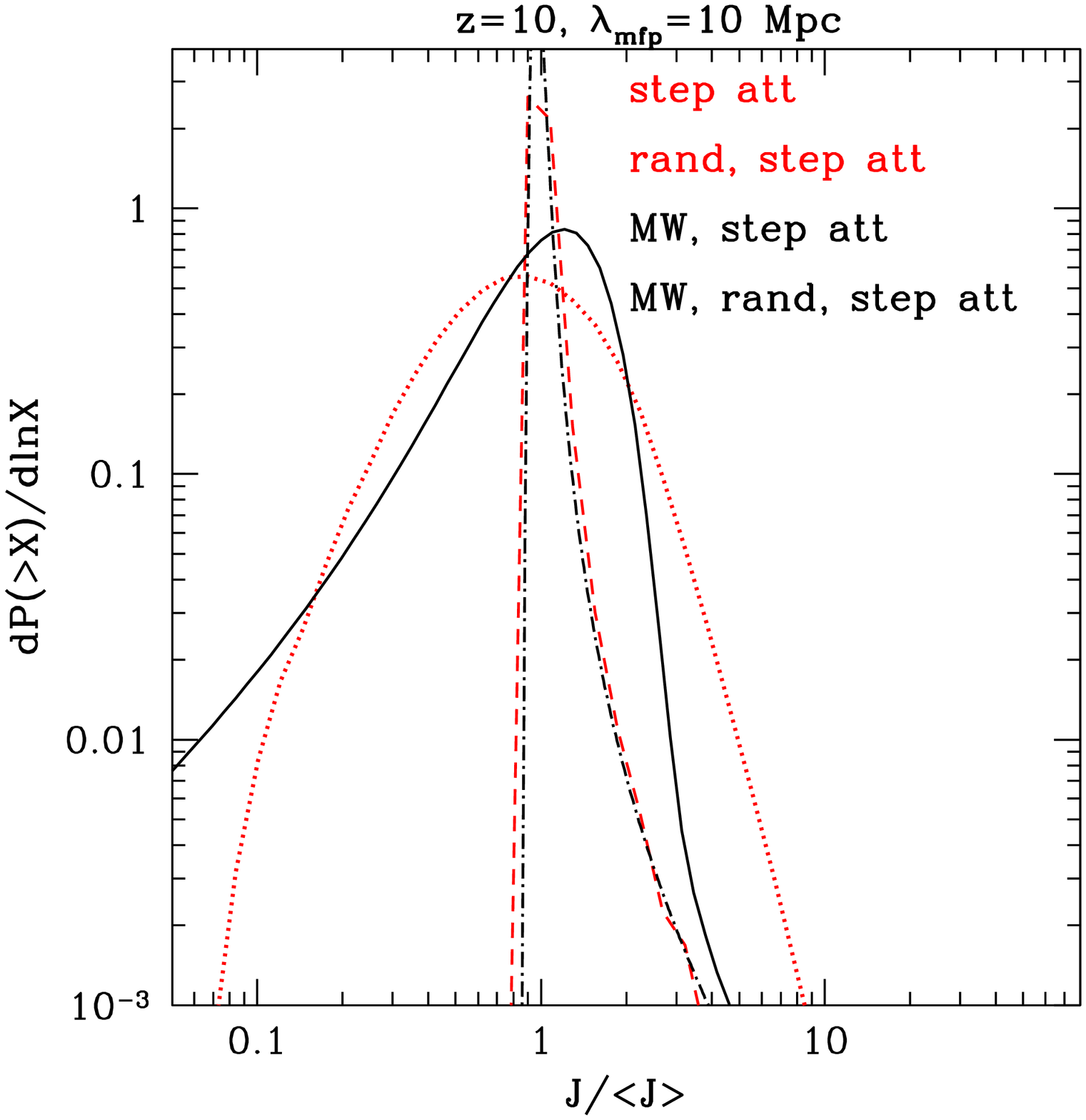}{3.3}{0.5}{.}{0.}
\caption{
Flux PDFs at $z=10$, assuming $\lmfp=10$ Mpc, $\Mmin=10^8 \Msun$ and DC=1.0.  The dotted curve was generated assuming a step-function attenuation, essentially $\tau=0$ for distances $r<\lmfp$ and $\tau=\infty$ at distances $r>\lmfp$.  The dashed curve is obtained in the same manner, but with randomized source locations (i.e. ignoring clustering).  The solid and dot-dashed curves correspond to the analytical model described in the text, including and ignoring clustering on scales $> \lmfp$ as described in equation~(\ref{eq:jdistbn_cluster}), respectively.
\label{fig:cluster}}
\vspace{-1\baselineskip}
\end{figure}

The clustering of sources broadens the flux PDF and creates a high-value tail in the distribution (e.g. \citealt{MD08, Dijkstra08}).  How strong is this effect and how much do the properties of the underlying sources affect these trends?

In Fig. \ref{fig:cluster}, we plot flux PDFs at $z=10$, assuming $\lmfp=10$ Mpc.
  The dotted curve in the figure was generated from the semi-numeric calculation assuming a step-function attenuation at $\lmfp=10$ Mpc, shown to be an adequate approximation in Fig. \ref{fig:mfp_att}.  The dashed curve is obtained in the same manner, but with randomized source locations.  We see that ignoring source clustering can severely underestimate the widths of the flux PDF.  Even the shape of the curves is quite different.

Fig.~\ref{fig:cluster} also compares the PDF to our analytic model.  The dot-dashed line uses the analog of  equation~(\ref{eq:jdist_r0}) but for step-wise attenuation (see \citealt{Zuo92} and \citealt{Furlanetto09}), again assuming Poisson distributed sources.  It closely matches the semi-numeric results for the same assumptions.  

%AM: moved sentance to below
The solid line includes linear clustering over the scale $\lambda_{\rm mfp}$ as described in equation~(\ref{eq:jdistbn_cluster}).   The clustering length of $M \sim \Mmin \sim 10^8 \Msun$ halos is $\sim$ 2 Mpc at this redshift, so the clustering on scales larger than the m.f.p. should be well into the linear regime and accurately predicted by the analytic model. Obviously, this simple prescription provides a relatively poor match to the full semi-numeric results:  although the large-scale clustering does broaden the distribution by a comparable amount, the shapes still disagree.  

Part of this difference is easy to understand and is a consequence of the analytic model's simple prescription for clustering:  the model not only fixes the \emph{local} attenuation volume to have a given density, but it also effectively assumes that surrounding regions have the same number of sources.  This artificially suppresses the contribution from distant galaxies (which provide 2/3 of the mean ionizing background in these models).  In practice, points inside low density regions should be exposed to higher flux values from neighboring higher density regions.  Such long-wavelength effects are missed in the simple analytic treatment, resulting in the spuriously long small-value tail in the PDF.

However, by the same logic our prescription places high-density regions near high-density neighbors, so one might expect us to overestimate the extent of the variation at high $J$ as well -- but in fact the prescription \emph{underestimates} the width there as well.  This is because, in contrast to the low-$J$ points, the radiation field at such high levels is not dominated by the background of distant sources but by nearby clusters of objects, for which smaller-scale clustering and nonlinear effects are important.

This is not particularly surprising in light of the analogy we have drawn to the halo model:  although that simple prescription does a remarkably good job of matching the dark matter power spectrum, it cannot easily be used to predict the actual density distribution of the IGM.  Indeed, \citet{Scherrer91} have shown that the PDF of the fractional deviation of $J$ from the mean value, $1+ \delta_J = J/\langle J \rangle$, is
\begin{equation}
{d p(>\delta_J) \over d\delta_J} = \int_{-\infty}^{\infty} {d s \over 2 \pi} e^{-i \delta_J s} 
\exp \left[ \sum_{p=1}^\infty {(i s)^p \over p!} \xi_J^{(p)}(0,0,...,0) \right],
\label{eq:pdf_scherrer}
\end{equation}
where $\xi_J^{(p)}({\bf r}_1,{\bf r}_2,...,{\bf r}_p)$ is the $p$-point correlation function of the radiation field.  Thus the PDF depends on \emph{all} higher order correlations at zero-lag, which are evidently strongly affected by the nonlinear clustering on small scales.  The simple addition of large-scale linear clustering is not sufficient to match the semi-numeric results.

\begin{figure}
\vspace{+0\baselineskip}
\myputfigure{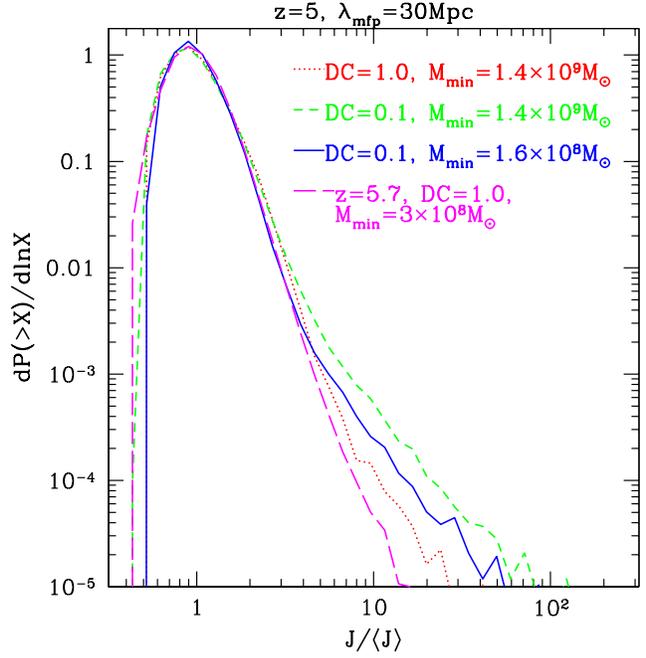}{3.3}{0.5}{.}{0.}
\caption{
Flux PDFs assuming $\lmfp=30$ Mpc. Dotted, short-dashed and solid curves correspond to $z=5$ and (DC, $\Mmin/\Msun$) = (1.0, $1.4\times10^9$), (0.1, $1.4\times10^9$), and (0.1, $1.6\times10^8$), respectively.   The long-dashed curve is generated from the $z=5.71$ star-formation rate output from the numerical results of \citet{TCL08}, as described in the text.
\label{fig:dc_PDFs}}
\vspace{-1\baselineskip}
\end{figure}

To see the impact of the source properties on the clustering signature in the flux PDF, in Fig. \ref{fig:dc_PDFs}, we plot several flux PDFs assuming $\lmfp=30$ Mpc.  The dotted, short-dashed and solid curves were all generated at $z=5$, but assuming various values of DC and $\Mmin$.  These two parameters drive the correlation function of sources and are crucial in constraining the UV luminosity to halo mass relation at high-$z$ (e.g. \citealt{Lee08}).  From Fig. \ref{fig:dc_PDFs}, we see that the low-end cut-off is generally unaffected by these parameters.  This is to be expected since, as we mention above, the low-end tail of the flux distribution corresponds to voids distant from UV sources, and are sensitive to the effective horizon $\lmfp$ (which is identical in all these curves).  Thus, most of the clustering imprint is in the high-value tails of the PDFs.

The figure also confirms that the widest flux PDF results from the model in which sources are the most biased and rare: having the highest $\Mmin$ and lowest DC.  As the duty cycle is increased, the high-value tail retracts.  A similar trend is seen as one lowers $\Mmin$.

 As mentioned above, the choice of $\Mmin=1.4\times10^9 \Msun$ was motivated by the fact that (DC, $\Mmin/\Msun$) = (1.0, $1.4\times10^9$) and (0.1, $1.6\times10^8$) result in the same number density of sources, and thus are useful in isolating the imprint of each these two parameters.  Even in our simple model where the duty cycle is not a function of mass, these corresponding curves are quite similar.  Thus it is unlikely that the flux PDF, even if it could be measured, could provide a useful constraint on these parameters.

The long-dashed curve in Fig. \ref{fig:dc_PDFs} shows the flux PDF obtained from the $z=5.71$ star-formation rate output from the hydrodynamical simulation of \citet{TCL08}, as described in \S \ref{sec:num}.  Despite differences in the star-formation prescription, this curve agrees well with the above trends.  Note that the slight offset at low $J$ is attributable to the fact that these fluxes were calculated including sources within $1.5 \lmfp$, instead of within $2 \lmfp$, as was the case for the semi-numerical boxes.

\section{The Flux Power Spectrum}
\label{sec:ps}

\subsection{Analytic Results}

\begin{figure}
\vspace{+0\baselineskip}
\myputfigure{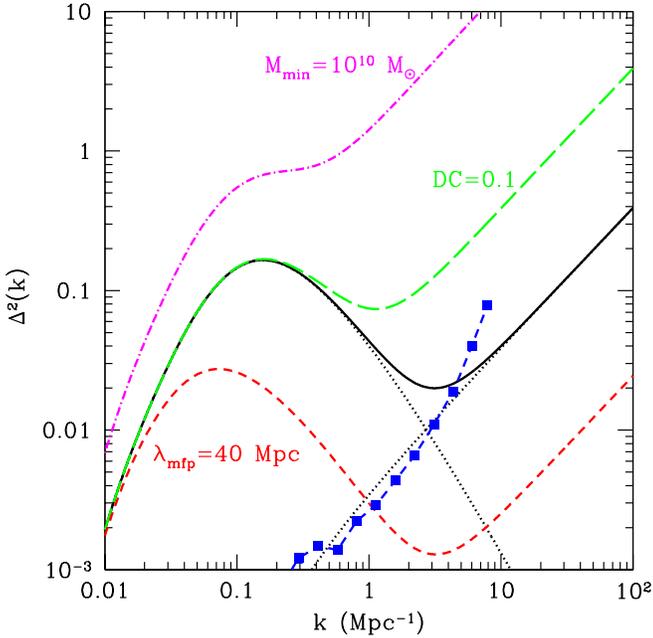}{3.3}{0.5}{.}{0.}
\caption{
Flux power spectra at $z=10$ computed with our analytic model.  The solid black line takes $\Mmin =10^8 \Msun$, DC$=1$, and $\lmfp=10$~Mpc; the dotted curves show the corresponding one- and two-source contributions.  The curve with solid squares shows the corresponding one-source power spectrum computed from the semi-numeric simulations.  The long-dashed line takes DC$=0.1$ instead, and the short-dashed line takes $\lambda_{\rm mfp}=40$~Mpc. The dot-dashed line takes $\Mmin =  10^{10} \Msun$.
\label{fig:analytic_pk}}
\vspace{-1\baselineskip}
\end{figure}

We begin our discussion of the flux power spectrum by examining the ``halo model" analytic results (which we expect to be more successful than the PDF).  Figure~\ref{fig:analytic_pk} shows some examples at $z=10$; here $\Delta^2(k, z) = k^3/(2\pi^2) P(k, z)$.  The solid curve takes $\Mmin =10^8 \Msun$, DC$=1$, and $\lmfp=10$~Mpc; the dotted curves show the corresponding one- and two-source contributions.  Clearly these two terms are cleanly separated in scale for this set of parameters.  The two-source term dominates on large scales; for $k \lambda_{\rm mfp} \ll 1$, it traces the linear power spectrum with an amplitude determined by the mean galaxy bias.  It is then damped at $k \lambda_{\rm mfp} \sim 1$:  the short-dashed curve, which takes $\lambda_{\rm mfp}=40$~Mpc shows how the turnover scale shifts with the attenuation length.

On small scales, the one-source term dominates:  it has $\Delta^2_{1s} \propto k^3 u_J(k)^2 \propto k/\lambda_{\rm mfp}^2$.  Thus its amplitude decreases rapidly with the attenuation length -- although this dependence is hard to separate from the dependence on overall source density.  The dashed curve with solid squares shows the one-source term computed from the semi-numeric simulations (with sources randomly distributed, so that the two-source term disappears).  It provides an excellent match to the analytic model, except on scales near $k \sim 10$~ Mpc$^{-1}$, where the discrete grid cells affect the semi-numeric power spectrum.

The dot-dashed curve takes a larger $\Mmin$.  Note that the power spectrum increases with the halo bias on very 
large scales. Also, the decreased source number density strongly increases the importance of the one-source term, giving a much less clean separation between the one-source and two-source regime.

\subsection{Numerical Results}

\begin{figure}
\vspace{+0\baselineskip}
\myputfigure{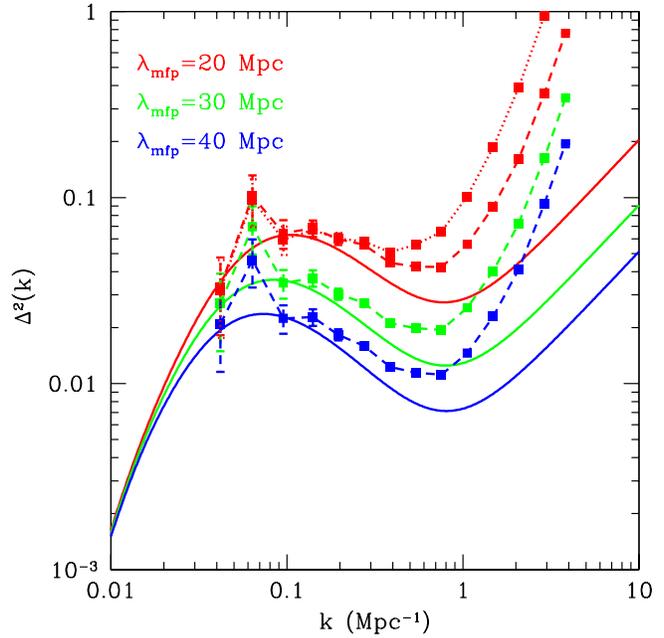}{3.3}{0.5}{.}{0.}
%\myputfigure{mfp_power_spec.eps}{3.3}{0.5}{.}{0.}
\caption{
Flux power spectra at $z=6$ from the analytic model ({\it solid curves}) and the semi-numeric simulations ({\it dashed curves}, with Poissonian error bars).  All curves assume $\Mmin=1.6\times10^8 \Msun$ and DC=0.; the three sets correspond to values of $\lmfp=$ 20, 30, 40 Mpc, from top to bottom.  The dotted curve shows the semi-numeric result at $z=5$ for $\lmfp=20$~Mpc.
\label{fig:mfp_ps}}
\vspace{-1\baselineskip}
\end{figure}

We now compare these analytic predictions to the semi-numerical simulations.  For the numerical calculations, we use the convention $\Delta^2(k, z) = k^3/(2\pi^2 V) ~ \langle|\delta_J({\bf k}, z)|^2\rangle_k$.  In Fig. \ref{fig:mfp_ps}, we plot the flux power spectra at $z=6$ for both the analytic ({\it solid curves}) and semi-numeric ({\it dashed curves}) models, generated assuming $\Mmin=1.6\times10^8 \Msun$ and DC=0.1.  Curves correspond to values of $\lmfp=$ 20, 30, 40 Mpc, from top to bottom.  

The agreement between the two calculations is quite impressive, especially at large scales.  The analytic model clearly shows the separation between the two-source and single-source terms; with the semi-numeric results the difference is less clear because the boxes are not large enough to sample the regime $k \lmfp \gg 1$, where $\Delta^2$ traces the matter power spectrum.  However, the amplitudes of the analytic and semi-numeric calculations match reasonably well on these scales:  the differences in the two-source regime are attributable to slightly different mean halo biases in the analytic and semi-numeric calculations.

However, the semi-numeric results have a steeper slope for the one-source component; this is most likely due to nonlinear clustering, which will steepen the correlation function on small scales, since Fig.~\ref{fig:analytic_pk} shows that the two different methods agree very well for the one-source term when clustering is ignored.  Interestingly, however, the crossover scale between the one-source and two-source regimes is accurately predicted by the analytic model.

The three sets of curves in Fig.~\ref{fig:mfp_ps} show that the m.f.p. has little impact in the shape of the spectrum over the scales probed by the semi-numeric simulations:  it primarily serves to decrease power fairly evenly on all scales with $k \lmfp \la 1$.  As described above, for smaller wave-numbers the attenuation length has no effect:  effectively, a larger attenuation length causes more damping relative to the linear power spectrum.

Finally, the dashed curve shows the flux power spectrum from the semi-numeric simulations at $z=5$ and $\lmfp=20$~Mpc.  By comparing Fig. \ref{fig:mfp_ps} and Fig. \ref{fig:mfp_pdfs}, we see that the flux power spectra are more sensitive to redshift than the flux PDFs.  The differences are similar for other values of $\lmfp$, and in the analytic model:  in all cases, the amplitude of the one-source component is somewhat larger at the lower redshift.  

This is easily understood from the analytic formalism.  The two source term is proportional to $\bar{b}^2 P_{\rm lin}(k,z)$; each of these terms evolves with redshift, but in the product the evolution nearly cancels at high redshift \citep{Oh01}.  The one source term is proportional to the second moment of the mass function (when integrated over mass -- eq.~\ref{eq:onesource}), which increases toward lower redshifts as more and more massive halos are assembled.

\begin{figure}
\vspace{+0\baselineskip}
\myputfigure{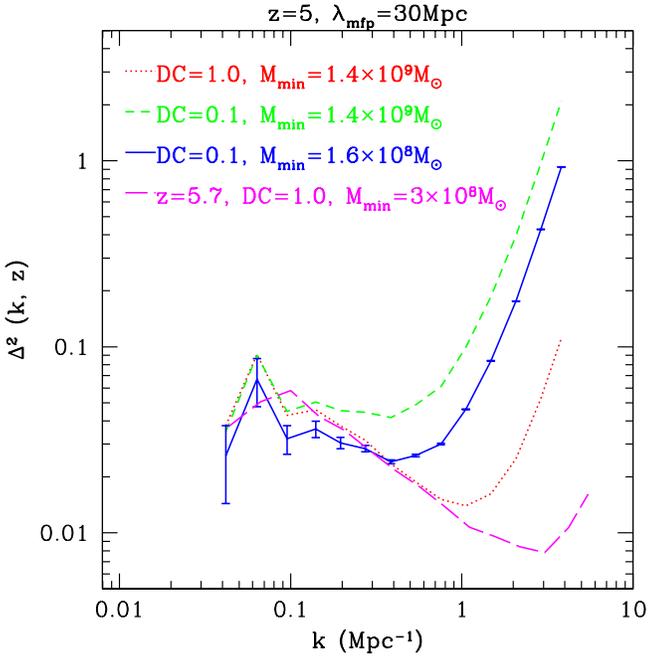}{3.3}{0.5}{.}{0.}
\caption{
Flux power spectra analogous to the models shown in Fig. \ref{fig:dc_PDFs}.
\label{fig:dc_ps}}
\vspace{-1\baselineskip}
\end{figure}

A similar argument holds for the power spectra presented in Fig. \ref{fig:dc_ps}, where we vary the DC and $\Mmin$ parameters.  The curves correspond to the same models presented in Fig. \ref{fig:dc_PDFs}.  Here we see drastic differences on moderate to small scales, because models in which the sources are rare (i.e. with low DC and high $\Mmin$) transition to the single-source regime on larger scales.  If observed, this statistic could be a powerful tool in constraining these parameters.  The flux field has the benefit of encoding information about ionizing sources which are too faint to be detected by galaxy surveys.  Unfortunately, such flux power spectra will be nearly impossible to extract from the Ly$\alpha$ forest, even given a very large number of sightlines (see Fig. \ref{fig:lim_tau_dc_ps} and associated discussion).

\section{Lyman $\alpha$ Forest}
\label{sec:forest}

The most obvious application of our flux field analysis is in the interpretation of the \lya\ forrest spectra following reionization.  Thus far, almost all estimates of the photoionization rate per hydrogen atom, $\Gamma_{12}\equiv\Gamma/10^{-12}$ s$^{-1}$, obtained from the \lya\ forest have assumed a uniform UV background (e.g. \citealt{Fan06}).  However, in \S \ref{sec:PDF} we showed that UV background actually has sizable fluctuations in this regime.  These fluctuations could potentially introduce bias and stocasticity in the determinations of $\Gamma_{12}$ from \lya\ forest attenuation measurements.

\citet{BH07} briefly addressed this issue by including a similar UV background model to our semi-numeric approach.  They confirm the results of lower redshift studies \citep{GH02, MW04, Bolton06} that assuming a homogeneous UVB underestimates the mean ionization rate obtained from \lya\ forest absorption, $\langle\Gamma_{12}\rangle$.  They find that this effect is of order $\sim$1\% -- 10\%, depending on $\lmfp$ and $\Mmin$ (see their Fig. 3).  However, for this analysis they used a single realization of a 30 $h^{-1}$ Mpc simulation box.  This box size is comparable to a single ionizing photon m.f.p at these redshifts \citep{Storrie-Lombardi94, MHR99, Miralda-Escude03, Faucher-Giguere08} and is several times smaller than the length of the GP troughs used to estimate $\langle\Gamma_{12}\rangle$.  Thus, their analysis was unable to fully take into account cosmic variance.  

Using our large simulation boxes, we wish to systematically investigate this effect.  As a first step, we do not create detailed mock absorption spectra.  Instead, for a given flux field, we merely calculate the \lya\ optical depth, $\tau_\alpha$, for each cell, assuming ionization equilibrium.  Then we average the flux decrements over the entire box to obtain a corresponding effective GP optical depth: $\taueffa \equiv -\ln \langle \exp[-\tau_\alpha] \rangle_{\bf x}$.

\begin{figure}
\vspace{+0\baselineskip}
\myputfigure{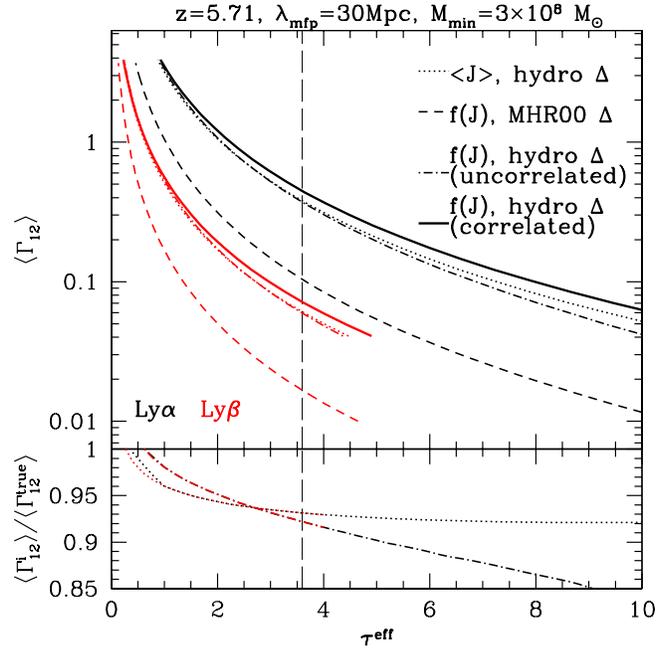}{3.3}{0.5}{.}{0.}
\caption{
{\it Top panel:} Mean ionization rate corresponding to an effective \lya\ ({\it black curves}) and \lyb\ ({\it red curves}) GP optical depth, for the $z=5.71$ hydrodynamical simulation box of \citet{TCL08}, assuming $\lmfp=30$ Mpc.  Dotted curves use the density field from the simulation, but assume a uniform UVB.  Dashed curves assume a fluctuating UVB from our simulations, but convolved with the \citet{MHR00} density distribution.  Dot-dashed curves use the fluctuating UVB and density field from the simulations, but randomly correlate the two.  Finally, solid curves use the correlated fluctuating UVB field and underlying density field from the simulations.
{\it Bottom panel:} Ratio of the mean ionization rate for each model $i$ and the ``true'' model shown as the solid curves in the top panel.  The vertical dashed line demarcates the average value of $\taueffa \sim 3.6$, roughly corresponding to the value measured from the SDSS QSOs at this redshift (taken from \citealt{Fan06}).
\label{fig:tau_to_gamma}}
\vspace{-1\baselineskip}
\end{figure}

In the top panel of Fig. \ref{fig:tau_to_gamma} we plot $\langle\Gamma_{12}\rangle$ for each corresponding effective \lya\ ({\it black curves}) and \lyb\ ({\it red curves}) GP optical depth, for the $z=5.71$ hydrodynamical simulation box, assuming $\lmfp=30$ Mpc.  The cell size for this simulation is 0.74 Mpc.  As mentioned above, the density field was calculated on a much finer grid of 0.19 Mpc, which resolves the Jeans length in the mean density, ionized IGM by a factor of few. The smoothed cell-size of 0.74 Mpc corresponds to the Jeans mass in the mean density ionized IGM, and is roughly comparable to the Doppler width of the \lya\ line at $T=10^4$ K, as well as to the Keck ESI resolution at this redshift.  Thus we expect our $\taueff$ $\leftrightarrow$ $\meanG$ mapping to be adequate [indeed, our results are consistent with the mock spectra analysis of \citet{BH07}], but we stress that we did not create detailed mock spectra.

The curves in Fig. \ref{fig:tau_to_gamma} show the effects of various simplifications on the value of $\langle\Gamma_{12}\rangle$ inferred from the $\taueff$ measured in the spectra.  The most accurate model, using both our fluctuating flux fields and underlying density fields, is shown by the solid curves.  By comparing the solid and dotted curves (or their ratios in the bottom panel), we see that assuming a uniform UVB makes very little impact on the estimated mean ionization rate per hydrogen atom.  At low values of $\taueff$, the two curves are practically indistinguishable, with the difference increasing with $\taueff$.  At $\taueffa \sim 3.6$ (roughly corresponding to the observed value at this redshift, e.g. \citealt{Fan06}), assuming a uniform UVB results in a $\sim5$\% underestimate in the value of $\langle\Gamma_{12}\rangle$.

It is fairly easy to interpret this behavior by looking back at Fig. \ref{fig:mfp_pdfs}.  The optical depth approximately scales as $\tau \propto \Delta^2 / \Gamma$, where $\Delta \equiv \rho/\bar{\rho}$ is the gas density in units of the mean.  Thus the optical depth distribution can be approximately obtained by convolving these two quantities.  From Fig. \ref{fig:mfp_pdfs}, we see that the distribution of $\Delta^2$ is much wider than all of the flux distributions.  Thus a uniform UVB, which treats the flux distributions as a delta function, is an accurate assumption.  Because of the exponential attenuation involved, the $\taueff$ statistic is sensitive to the low-end tail of the density distribution.  As the ionizing background gets fainter (i.e. increasing $\taueff$), the $\taueff$ statistic is sensitive to an increasingly narrow range in the low-end tail of the optical depth distribution.  It is in this regime that the non-zero width of the ionizing background PDF becomes evident, as the convolution with $\Delta^2$ is mostly evident in smearing-out this low-end tail.  Thus assuming a uniform UVB underestimates the true $\meanG$, but only slightly, with this underestimate increasing with increasing $\taueff$.

Figure \ref{fig:tau_to_gamma} also shows the impact of other common assumptions in the $\taueff$ $\leftrightarrow$ $\meanG$ mapping.  The dot-dashed curve is generated by using the same UVB and density fields as in the solid curve, but randomly pairing the two.  This also results in a slight underestimate of $\meanG$, though a little more notable than the uniform UVB assumption.  Since underdense pixels are preferentially farther away from most sources, where the UVB is lower (c.f. Fig. \ref{fig:pics}), not correlating the two fields results in underdense pixels receiving fluxes which are too strong, thus lowering $\meanG$.  Nevertheless, the effect is still small because of the relative width of the PDFs argument in the above paragraph.

%Am: added at end
Finally, the most drastic effect, resulting in an underestimate of a factor of $\sim$2 in our models, comes from using the analytic density PDF from \citet{MHR00}.  
This offset
%, which appears in all studies that use the \citet{MHR00} distribution,
 is normally attributed to errors in the fluctuating Gunn-Peterson approximation, which ignores effects such as the line wings, peculiar velocities, and blending.  However, our numeric results make a similar approximation, so that cannot be the source of the error here.  Instead, it may be differences between the assumed density distribution (generated to fit numerical simulations at redshifts $z\sim$ 2--4, using a simulation box of only 10 $h^{-1}$ Mpc) and that of the simulation.  At higher redshifts (i.e. $\taueff$ values), the transmission becomes dominated by rare voids \citep{OF05, LOF06}, and so cosmic variance is very important.  Proper analysis requires large simulation boxes with accurate hydrodynamical treatment and direct higher-redshift outputs.  However, if the semi-analytic method is renormalized using a simulation (as is typical), the \emph{shapes} of the curves are similar enough that the errors will be relatively small, so the approximation is still useful for understanding the evolution of $\meanG$.  Furthermore, it is not clear how this offset is affected by the assumed thermal history of the gas, which sets the effective local smoothing scale.  A detailed exploration of this is beyond the scope of this work, and merits its own study.

\begin{figure}
\vspace{+0\baselineskip}
\myputfigure{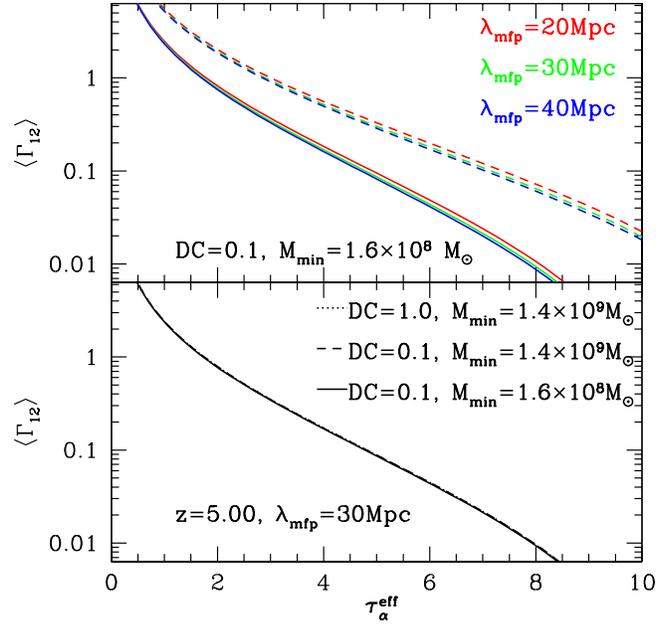}{3.3}{0.5}{.}{0.}
\caption{
$\taueff$ $\leftrightarrow$ $\meanG$ mappings using our semi-numerical simulations, assuming various flux models.  In the top panel, solid curves correspond to $z=5$, while dashed curves correspond to $z=6$.
\label{fig:mfp_dc_tau_to_gamma}}
\vspace{-1\baselineskip}
\end{figure}

In Fig. \ref{fig:mfp_dc_tau_to_gamma}, we show the $\taueff$ $\leftrightarrow$ $\meanG$ mappings using our semi-numeric simulations, assuming various flux models.  Note that the semi-numerical density fields slightly overpredict the presence of very rare voids which dominate this statistic (Mesinger et al., in preparation), thereby over-predicting the transmission for a given UVB.  It is remarkable how insensitive this mapping is to the details of the UVB, despite the notable differences in the flux PDFs and power spectra shown in the previous section.  Again, this is due to the fact that the fluctuations in the UVB, and the differences among various models of the UVB, are overshadowed by the fluctuations in the density field.  The only notable differences among the curves is due to the density field evolution from $z=6$ to $z=5$.

We note that our results are in general agreement with \citet{BH07}, who found that assuming a homogeneous UVB underestimates $\langle\Gamma_{12}\rangle$, by a few percent over this range of $\lmfp$ values.  However, they find a somewhat stronger dependence on $\lmfp$ then we do.  This is most likely a result of cosmic variance attributable to the small simulation box they used to study this effect.

\begin{figure}
\vspace{+0\baselineskip}
\myputfigure{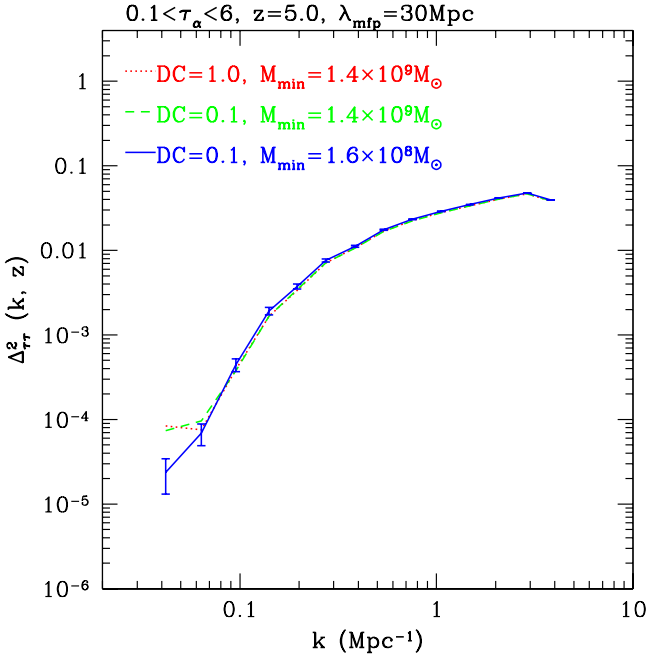}{3.3}{0.5}{.}{0.}
\caption{
Optical depth power spectra at $z=5$ and $\lmfp=30$ Mpc, within the range 0.1$<\tau_\alpha<$6, approximately corresponding to the range probed by current instruments.
\label{fig:lim_tau_dc_ps}}
\vspace{-1\baselineskip}
\end{figure}

Finally, we wish to quantify how easy is it to distinguish various UVB models from observations of the Ly$\alpha$ forest.  One can see from Fig. \ref{fig:dc_ps} that the shape of the flux power spectrum encodes information about the clustering properties of the sources.  As mentioned above, the flux field (as potentially extracted from the Ly$\alpha$ forest), has a great benefit in being able to probe the impact of ionizing sources below the detection thresholds of galaxy surveys.  Hence, in Fig. \ref{fig:lim_tau_dc_ps} we show the optical depth power spectra at $z=5$ and $\lmfp=30$ Mpc for some of the same models as in Fig. \ref{fig:dc_ps}.  Unfortunately, the optical depth power spectra of these models are practically indistinguishable from each other, despite the large differences shown in Fig. \ref{fig:dc_ps}.  This result could have been foreshadowed from Fig. \ref{fig:mfp_dc_tau_to_gamma} and is again due to the fact that the density field and its fluctuations swamp the UVB signal.

However, we note that the density fluctuations decrease as redshift increases, while fluctuations in the UVB increase.  The UVB becomes more inhomogeneous not only through the evolution of $\lmfp$ and the bias of sources, but it is also modulated by the ionized bubbles throughout reionization (whenever that epoch may be).  Thus inhomogeneities in the flux field become increasingly important at higher redshifts,
although whether this can be observed in the highly saturated forest at $z \ga 6$, where the density fluctuations themselves cause substantial fluctuations in $\taueff$ \citep{LOF06}, is questionable.

\section{Conclusions}
\label{sec:conc}

We systematically study the expected fluctuations in the hydrogen ionizing background in the epoch following reionization ($z\sim$ 5--6).  Unlike at lower redshifts ($z\lsim4$), the UVB can be quite inhomogenous in this regime.  This is due to the smaller mean free path of ionizing photons, as well as the clustering of increasingly rare ionizing sources.

We confirm that there is a sizable spread (the widths of the PDFs at half of the maximum likelihoods spanning factors of $\sim$ 2--4) and asymmetry in the distribution of ionizing fluxes in a cosmological volume.  Expected values of the m.f.p. are large enough ($\lmfp\gsim20$ Mpc), and sources are ubiquitous enough, that the PDF of the ionizing background is asymmetric, with a tail extending to higher values.  This high-value tail is set by clustering on small scales, and is insensitive to the contributions of distant sources (i.e. $\lmfp$) \citep{MD08}.  We also find that the PDFs are most sensitive to the m.f.p., with higher values of $\lmfp$ increasingly truncating the low-value tails. Furthermore, we note that the analytic formalism of \citet{MW04} severely underestimates the width of the flux PDF at this epoch, since it ignores the clustering of sources.

The power spectra of the ionizing background are even more sensitive to the details of the source model.  We have shown that an approach analogous to the halo model, in which contributions from the proximity zones of single sources (which dominate fluctuations on small scales and are independent of the large-scale galaxy distribution) and multiple sources (which dominate on large scales, and which depend on the large-scale clustering of the ionizing sources) are separated, provides a good match to the simulated results, although nonlinear clustering provides even more small-scale power than the analytic model predicts.  The single-source component is particularly sensitive to the number density of galaxies; unfortunately, these power spectra will be nearly impossible to observe with the \lya \ forest.

We also model the imprint of various UVBs in the \lya\ forest at $z\sim$ 5--6.  We find that the \lya\ forest spectra are extremely insensitive to the details of the UVB, despite marked differences in the PDFs and power spectra of our various UVBs.  This is attributable to the fact that the \lya\ optical depth scales as $\tau\propto\Delta^2/\Gamma$, and the squared overdensity distribution, $f(\Delta^2)$, is much wider than all of our $\Gamma$ distributions.  In fact, even the extreme scenario of a uniform background (i.e. a delta function $\Gamma$ distribution), only underestimates the value of $\langle\Gamma_{12}\rangle$ inferred from the \lya\ forrest by a few percent.  Thus our results justify the common assumption of a uniform UVB in \lya\ forest analysis.

We also confirm that accurate modeling of the density field is very important in the $\taueff \leftrightarrow \langle\Gamma_{12}\rangle$ relation.  We find significant differences (factor of $\sim$ 2) in this relation among the numerical and extrapolated \citet{MHR00} density distributions (though we stress again that we did not create detailed mock spectra in this work). At higher redshifts, the $\taueff$ statistic increasingly depends on a narrower range in the low-value density PDF.  Large-scale hydrodynamical modeling of the statistics of these rare voids is therefore essential in interpreting the high-redshift \lya\ forest.  \citet{Pawlik09} have recently studied the IGM density field during reionization, and do find significant differences from the \citet{MHR00} fitting formula, although their simulation boxes are much too small to study the rare voids pertinent to these efforts.

An interesting comparison can be made between our analysis of hydrogen reionization and similar effects during and after helium reionization (Dixon \& Furlanetto, in preparation).  In the latter case, the PDF of $J$ is much broader (spanning over an order of magnitude), and the fluctuations have a much larger effect on the HeII \lya\ forest, introducing a factor $\sim 2$ bias in estimates of $\langle \Gamma \rangle$ when the background is assumed uniform.  This has important implications for the HeII \lya\ forest during reionization:  unlike the case we have studied here, the large fluctuations in the ionizing background lead to substantial opacity gaps even during the middle phases of helium reionization.

\vskip+0.5in

We thank Hy Trac and Renyue Cen for the use of the $z=5.71$ density and source output of their simulations, as well as for stimulating conversations. We also thank James Bolton for helpful comments on the manuscript. Support for this work was provided by NASA through Hubble Fellowship grant \#HF-01222.01 to AM, awarded by the Space Telescope Science Institute, which is operated by the Association of Universities for Research in Astronomy, Inc., for NASA, under contract NAS 5-26555.  SF was partially supported by the NSF through grant AST-0607470 and by the David and Lucile Packard Foundation.

\bibliographystyle{mn2e}
\bibliography{ms}

\end{document}

%% file: defns.tex
\newcommand{\lya}{Ly$\alpha$}
\newcommand{\lyb}{Ly$\beta$}

\newcommand{\hs}{\hspace{1mm}}

\newcommand{\Msun}{M_\odot}

\newcommand{\Mmin}{M_{\rm min}}

\newcommand{\lmfp}{\lambda_{\rm mfp}}

\newcommand{\flux}{f({\bf x}, z)}

\newcommand{\apjl}{ApJ}
\newcommand{\apjs}{ApJS}
\newcommand{\aap}{A\&A}